\documentclass[12pt,a4paper]{article}
\usepackage{amsmath,amssymb,amsthm}
\usepackage[margin=1.0in]{geometry}
\usepackage{cite}
\usepackage[colorlinks=true
,urlcolor=blue
,anchorcolor=blue
,citecolor=blue
,filecolor=blue
,linkcolor=blue
,menucolor=blue
,pagecolor=blue
,linktocpage=true
,pdfproducer=medialab
,pdfa=true
]{hyperref}
\numberwithin{equation}{section}

\def\k{\kappa}
\def\l{\lambda}
\def\m{\mu}
\def\n{\nu}

\def\r{\rho}

\def\be{\begin{equation}}
\def\ee{\end{equation}}
\def\bea{\begin{eqnarray}}
\def\eea{\end{eqnarray}}

\def\pa{\partial}

\def\lp{\left(}
\def\rp{\right)}

\def\ie{{\it i.e., }}

\makeatletter
\renewcommand\section{\@startsection {section}{1}{\z@}%
	{-3.5ex \@plus -1ex \@minus -.2ex}
	{2.3ex \@plus.2ex}%
	{\normalfont\large\bfseries}}
\renewcommand\subsection{\@startsection{subsection}{2}{\z@}%
	{-3.25ex\@plus -1ex \@minus -.2ex}%
	{1.5ex \@plus .2ex}%
	{\normalfont\bfseries}}
\makeatother

\begin{document}

\begin{center}
\addtolength{\baselineskip}{.5mm}
\thispagestyle{empty}
\begin{flushright}
\end{flushright}

\vspace{20mm}

{\Large  \bf Gauge field corrections to 11-dimensional supergravity via dimensional reduction}
\\[15mm]
{Hamid R. Bakhtiarizadeh\footnote{bakhtiarizadeh@sirjantech.ac.ir}}
\\[5mm]
{\it Department of Physics, Sirjan University of Technology, Sirjan, Iran}

\vspace{20mm}

{\bf  Abstract}
\end{center}
Using the fact that eleven-dimensional supergravity yields type IIA supergravity under dimensional reduction on a circle, we determine higher-derivative terms of eleven-dimensional supergravity including the $ R^4 $, $ ({\partial {F_4}})^2 R^2 $ and $ ({\partial {F_4}})^4 $ terms.

\vfill
\newpage


\section{Introduction}

The low-energy effective action of M-theory is known as the eleven-dimensional supergravity. This theory is described by massless modes of M-theory (the graviton, the three-form and the gravitino), which contains a membrane as a fundamental object. This theory also consists of the lowest-order supergravity action \cite{Cremmer:1978km} plus an infinite number of higher-derivative terms beyond the leading order. 

There exists a variety of methods which can be used to capture these higher-derivative terms. Let us briefly review some of them. The perturbative analyses of the scattering amplitudes is one of the important methods to determine the structure of the higher-derivative corrections to the  11d supergravity \cite{Deser:1998jz,Deser:2000xz}. Besides the approaches based on the perturbation analyses, one of the famous methods is the analysis performed by computing the scattering amplitudes of superparticles \cite{Green:1997as,Green:1997di,Russo:1997mk,Green:1999pu,Dasgupta:2000df,Plefka:2000gv,Anguelova:2004pg,Peeters:2005tb}, the superfield method \cite{Cremmer:1980ru,Green:1999by,Nicolai:2000ht,Green:2005ba,Cederwall:2004cg,deHaro:2002vk,Howe:2003cy,Plefka:1998xu,Peeters:2000qj} and by applying Noether's method \cite{Hyakutake:2006aq}. 

Among these approaches, we employ the straightforward dimensional reduction method to determine the higher-derivative corrections to 11d supergravity. We assume that all fields are independent of the coordinate $ z = x^{11} $ which we choose to correspond to a space-like direction ($ \eta^{(11)}_{zz} = 1 $) and then we rewrite the fields and action in a ten-dimensional form.

Let us now consider the dimensional reduction of the bosonic fields of 11d supergravity, the metric and the three-form \cite{Scherk:1979zr,Huq:1983im}. The dimensional reduction of the metric gives rise to the ten-dimensional metric, a vector field, and a scalar (the dilaton). According to this, the metric of eleven-dimensional theory has to be expressed in terms of the ten-dimensional one as follows:
\bea
g^{(11)}_{\m \n} = e^{-\frac{2}{3}\Phi}g_{\m \n}+e^{\frac{4}{3}\Phi} {C_1}_{\m}{C_1}_{\n},\qquad
g^{(11)}_{\m z} = e^{\frac{4}{3}\Phi} {C_1}_{\m}\qquad {\rm and} \qquad
g^{(11)}_{zz} = e^{\frac{4}{3}\Phi},\label{metric}
\eea
whereas the dimensional reduction of the 3-form potential in $ D = 11 $ gives rise to a three-form and a two-form which are the fields of the 10d supergravity theory
\bea
{C_3}^{(11)}_{\m \n \r}= {C_3}_{\m \n \r}\qquad {\rm and} \qquad {C_3}^{(11)}_{\m \n z}= B_{\m \n},
\eea
with the corresponding field strengths $ F_4=dC_3 $ and $ H=dB $ given by
\bea
{F_4}^{(11)}_{\m \n \r \l}= {F_4}_{\m \n \r \l}\qquad {\rm and} \qquad {F_4}^{(11)}_{\m \n \r z}= H_{\m \n \r}.\label{4form}
\eea

The terms we would like to obtain consist of 4-form field strength and Riemann tensor. The dimensional reduction of 4-form field strength is given by Eq. (\ref{4form}), whereas the dimensional reduction of the Riemann tensor needs more considerations.  

For our intended purposes, it is sufficient to study the dimensional reduction of 11-dimensional supergravity which involves four massless fields. So we need the transformations (\ref{metric}) at the linear order. Assuming that the massless fields are small perturbations around the flat background, \ie
\bea
g_{\m \n}=\eta_{\m \n}+2\k h_{\m \n};\;
  \Phi=\phi_0+2\k \phi;\;{C_1}_{\m}=2\k{c_1}_{\m}.
\eea
The transformation of $ g_{\m \n} $, which is introduced in (\ref{metric}), takes the following linear form for the perturbations:
\bea
h^{(11)}_{\m \n}=h_{\m \n}.\label{pert}
\eea
On the other hand, the linearized Riemann curvature is defined as
\bea
R_{\m \n}{}^{ \r \l}= \k \pa_{[\m} \pa^{[\r} h_{\n]}{}^{\l]}.
\eea
The Eq. (\ref{pert}) implies that the transformation of the linearized Riemann tensor, when carries no Killing index, is
\bea
R^{(11)}_{\m \n \r \l}=R_{\m \n \r \l}.\label{Rrule}
\eea

The requirement of the dimensional reduction is a powerful tool to restrict the form of an effective action. The procedure of the dimensional reduction method is well known and quite simple. First we prepare an ansatz for the higher derivative effective action in which each term has some unknown coefficients. Then we consider the dimensional reduction of the ansatz by splitting the eleven-dimensional indices into the ten-dimensional ones and the eleventh index $ z $. Some of the generated terms can be transformed to the known couplings in ten dimensions under dimensional reduction rules. The comparison of these terms gives rise simultaneous equations among the unknown coefficients in the ansatz. By solving these equations and substituting the solutions into the ansatz, one can determine the possible forms of the higher-derivative effective action.

The content of our paper is as follows. In Sec. \ref{R4}, we first construct an ansatz for $ R^4 $ terms with unknown coefficients in eleven dimensions and then derive them by forcing the ansatz to match with the known $ R^4 $ terms in ten dimensions. In Sec. \ref{F2R2}, we follow the same procedure to determine the $ (\pa {F_4})^2 R^2 $ terms in eleven dimensions. Finally, in Sec. \ref{F4} we will obtain $ (\pa {F_4})^4 $ terms. Sec. \ref{dis} is devoted to discussion.
 
\section{$ R^4 $ terms}\label{R4}

An ansatz for the higher derivative effective action, which includes quartic terms of the Riemann tensor \cite{Hyakutake:2006aq}, is parametrized by
\bea
&& C_1^{\text{}} R_{abcd} R^{abcd} R_{efgh} R^{efgh} + C_2^{\text{}} R_{abcd} R^{abc}{}_{e} R^{d}{}_{fgh} R^{efgh} \nonumber \\ 
&&+ C_3^{\text{}} R_{abcd} R^{ab}{}_{ef} R^{cd}{}_{gh} R^{efgh} + C_4^{\text{}} R_{abcd} R^{a}{}_{e}{}^{c}{}_{g} R^{b}{}_{f}{}^{d}{}_{h} R^{efgh} \nonumber \
\\ 
&& + C_5^{\text{}} R_{abce} R^{ab}{}_{dg} R^{c}{}_{f}{}^{d}{}_{h} R^{efgh} + C_6^{\text{}} R_{abce} R^{ab}{}_{df} R^{cd}{}_{gh} R^{efgh} \nonumber \\ 
&& + C_7^{\text{}} R_{abce} R^{a}{}_{d}{}^{c}{}_{g} R^{b}{}_{f}{}^{d}{}_{h} R^{efgh}.
\eea
Note that the terms which include the scalar curvature and Ricci tensor are removed by using the field redefinition. 

Upon dimensional reduction of our ansatz to ten dimensions, it should be possible to extract the terms in which the Riemann tensor carries no Killing index. These terms are transformed to ten-dimensional ones according to the rule (\ref{Rrule}). Then, we match the obtained results, which have the same structure as ansatz but with indices in ten dimensions, with the known $ R^4 $ terms computed a long time ago by Gross and Sloan \cite{Gross:1986mw}. This match of the dimensionally-reduced action provides strong consistency check on our computations and results in the following relations between the unknown coefficients\footnote{The calculations in this paper have been done with the xAct package of Mathematica\cite{Nutma:2013zea}.}:
\bea
\{C_2 \to -16 C_1, C_3 \to 2 C_1, C_4 \to 16 C_1, C_5 \to -32 C_1, C_6 \to -32 + 16 C_1, C_7 \to 128 - 32 C_1\}.\nonumber
\eea

Inserting these conditions into the ansatz leads to the following $ R^4 $ terms in $ 11 $ dimensions:
\bea
e^{-1}{\cal L}_{R^4}=32 \lp 4 R_{abce} R^{a}{}_{d}{}^{c}{}_{g} R^{b}{}_{f}{}^{d}{}_{h} R^{efgh} -  R_{abce} R^{ab}{}_{df} R^{cd}{}_{gh} R^{efgh} \rp,\label{RRRR}
\eea
plus some other terms with unknown coefficients which implicitly are zero. The reason is that they vanishes when we write them in terms of independent variables in which all symmetries (including mono- and multi-term symmetries), mass-shell and on-shell conditions as well as conservation of momentum are applied. In the above equation, $ e $ denotes $ \sqrt{-g} $, where $ g $ is the determinant of the metric in 11 dimensions. 

\section{$ (\pa {F_4})^2 R^2 $ terms}\label{F2R2}

Let us now consider the ansatz of the $ (\pa {F_4})^2 R^2 $ part. By imposing the linearised lowest-order equations of motion \cite{Peeters:2005tb}, one obtains 24 possible terms in the action
\bea
&& C_1^{\text{}} F^{agh}{}_{i}{}^{,e} F^{bdfi,c} R_{abcd} \
R_{efgh} + C_2^{\text{}} F^{acg}{}_{i}{}^{,e} F^{bdfi,h} R_{abcd} R_{efgh} \nonumber \\ 
&& + C_3^{\text{}} F^{acg}{}_{i}{}^{,e} F^{bdhi,f} R_{abcd} \
R_{efgh} + C_4^{\text{}} F^{cdgh}{}_{,i} F^{iabe,f} R_{abcd} R_{efgh} \nonumber \\ 
&&+ C_5^{\text{}} F^{bc}{}_{hi}{}^{,a} \
F^{fghi,e} R_{abcd} R_{efg}{}^{d}+ C_6^{\text{}} F^{be}{}_{hi}{}^{,a} F^{fghi,c} R_{abcd} \
R_{efg}{}^{d}\nonumber \\ 
&& + C_7^{\text{}} F^{be}{}_{hi}{}^{,a} \
F^{cfhi,g} R_{abcd} R_{efg}{}^{d}+ C_8^{\text{}} F^{ce}{}_{hi}{}^{,a} \
F^{fghi,b} R_{abcd} R_{efg}{}^{d}\nonumber \\ 
&& + C_9^{\text{}} F^{bghi,f} F^{ce}{}_{hi}{}^{,a} R_{abcd} \
R_{efg}{}^{d}+ C_{10}^{\text{}} F^{abfh,i} F^{ceg}{}_{h,i} R_{abcd} R_{efg}{}^{d} \nonumber \\ 
&&+ C_{11}^{\text{}} F^{abf}{}_{i,h} F^{cegh,i} R_{abcd} R_{efg}{}^{d} + C_{12}^{\text{}} F^{bahi,g} F^{ef}{}_{hi}{}^{,c} R_{abcd} R_{efg}{}^{d}  \nonumber \\ 
&&+ C_{13}^{\text{}} F^{bf}{}_{gh,i} F^{degh,i} R_{abcd} R_{e}{}^{a}{}_{f}{}^{c}   + C_{14}^{\text{}} F^{bd}{}_{gh,i} \
F^{efgh,i} R_{abcd} R_{e}{}^{a}{}_{f}{}^{c} \nonumber \\ 
&& + C_{15}^{\text{}} F^{dghi,e} F^{f}{}_{ghi}{}^{,b} R_{abcd} R_{e}{}^{a}{}_{f}{}^{c} + C_{16}^{\text{}} F^{d}{}_{ghi}{}^{,b} F^{fghi,e} R_{abcd} R_{e}{}^{a}{}_{f}{}^{c} \nonumber \\ 
&& + C_{17}^{\text{}} \
F^{e}{}_{ghi}{}^{,b} F^{fghi,d} R_{abcd} R_{e}{}^{a}{}_{f}{}^{c}+ C_{18}^{\text{}} F^{bghi,d} F^{ef}{}_{gh,i} R_{abcd} R_{e}{}^{a}{}_{f}{}^{c} \nonumber \\ 
&&+ C_{19}^{\text{}} \
F^{ae}{}_{gh,i} F^{bfgh,i} R_{abcd} R_{ef}{}^{cd}  + C_{20}^{\text{}} \
F^{e}{}_{ghi}{}^{,a} F^{fghi,b} R_{abcd} R_{ef}{}^{cd}\nonumber \\ 
&&+ C_{21}^{\text{}} F^{bghi,f} F^{e}{}_{ghi}{}^{,a} R_{abcd} R_{ef}{}^{cd}+ C_{22}^{\text{}} F_{fghi}{}^{,b} F^{fghi,e} R_{abcd} R_{e}{}^{acd} \nonumber \\ 
&&+ C_{23}^{\text{}} F^{b}{}_{fgh,i} F^{efgh,i} R_{abcd} R_{e}{}^{acd}+C_{24}^{\text{}} F^{eghi,f} F_{fghi,e} R_{abcd} R^{abcd},\label{FFRR} 
\eea
where comma on the 4-form indices refers to a partial derivative with respect to the index afterwards. To find the unknown coefficients, we impose the following two constraints on the above ansatz: 
\begin{enumerate}
	\item The terms with structure $ (\pa {F_4})^2 R^2 $ in $ D = 11 $ should transform to $ (\pa {F_4})^2 R^2 $ in $ D = 10 $ under dimensional reduction.
	\item Upon  dimensional reduction rules, the terms with structure $ (\pa {F_4}_{z})^2 R^2 $ in $ D = 11 $ should convert to $ (\pa H)^2 R^2 $ in $ D = 10 $.
\end{enumerate}

By splitting the indices of ansatz, one may consider the terms with structure $ (\pa {F_4})^2 R^2 $ in which the $ 4 $-form field strength and the Riemann tensor carry no Killing index. One can shift them to ten dimensions according to the rules (\ref{4form}) and (\ref{Rrule}). These terms are similar to the 11-dimensional ones but with indices in ten dimensions, as was expected. 

The corresponding couplings in type IIA supergravity have been previously found in \cite{Policastro:2006vt,Garousi:2013nfw,Bakhtiarizadeh:2017bpl,Bakhtiarizadeh:2013zia}. On the other hand, the terms $ (\pa H)^2 R^2 $ in ten dimensions which are obtained by applying the above second constraint on the terms in which each $ 4 $-form field strength carries one Killing index and the Riemann tensors carry no one, have the following form 
\bea
&&-2 C_6 H_{a}{}^{ef,g} H^{abc,d} R_{bdg}{}^{h} R_{cefh} -  C_1 H^{abc,d} H^{efg,h} R_{abde} R_{cfgh} \nonumber \\ 
&& -  C_{10} H^{abc,d} H^{efg}{}_{,d} R_{abe}{}^{h} R_{cfgh} - 2 C_7 H_{a}{}^{ef,g} 
H^{abc,d} R_{bde}{}^{h} R_{cfgh} \nonumber \\ 
&& + 2 C_{19} H_{a}{}^{ef}{}_{,d} H^{abc,d} R_{be}{}^{gh} R_{cfgh} + C_3 H^{abc,d} 
H^{efg,h} R_{aebf} R_{cgdh} \nonumber \\ 
&& + 2 C_9 H_{a}{}^{ef,g} H^{abc,d} R_{bge}{}^{h} 
R_{chdf} + C_2 H^{abc,d} H^{efg,h} R_{aebf} R_{chdg} \nonumber \\ 
&& + 2 C_{13} H_{a}{}^{ef}{}_{,d} H^{abc,d} R_{b}{}^{g}{}_{e}{}^{h} R_{chfg} + 3 C_{21} 
H_{ab}{}^{e,f} H^{abc,d} R_{cf}{}^{gh} R_{degh} \nonumber \\ 
&&+ 3 C_{20} H_{ab}{}^{e,f} H^{abc,d} R_{ce}{}^{gh} R_{dfgh} - 2 C_8 H_{a}{}^{ef,g} 
H^{abc,d} R_{bec}{}^{h} R_{dgfh} \nonumber \\ 
&& + 3 C_{17} H_{ab}{}^{e,f} H^{abc,d} R_{c}{}^{g}{}_{e}{}^{h} R_{dgfh} - 2 C_{12} 
H_{a}{}^{ef,g} H^{abc,d} R_{bcg}{}^{h} R_{dhef} \nonumber \\ 
&&+ 3 C_{15} H_{ab}{}^{e,f} H^{abc,d} R_{c}{}^{g}{}_{f}{}^{h} R_{dheg} - 3 C_{23} 
H_{ab}{}^{e}{}_{,d} H^{abc,d} R_{c}{}^{fgh} R_{efgh} \nonumber \\ 
&& - 4 C_{22} H_{abc}{}^{,e} H^{abc,d} R_{d}{}^{fgh} R_{efgh} + 2 C_5 H_{a}{}^{ef,g} 
H^{abc,d} R_{bdc}{}^{h} R_{egfh} \nonumber \\ 
&& - 2 C_{18} H_{ad}{}^{e,f} H^{abc,d} R_{b}{}^{g}{}_{c}{}^{h} R_{egfh} + 2 C_{14} 
H_{a}{}^{ef}{}_{,d} H^{abc,d} R_{b}{}^{g}{}_{c}{}^{h} R_{egfh} \nonumber \\ 
&& + 3 C_{16} H_{ab}{}^{e,f} H^{abc,d} R_{c}{}^{g}{}_{d}{}^{h} R_{egfh} + 3 C_{24} 
H_{abd,c} H^{abc,d} R_{efgh} R^{efgh}.
\eea

It also has been shown that the $ (\pa H)^2 R^2 $ terms in the 10-dimensional effective action can be obtained from the known $ R^4 $ action by extending the Riemann curvature to the generalized Riemann curvature \cite{Gross:1986mw}. 

By comparing the results obtained from the above two constraints with the corresponding ones in ten dimensions, one observes that both constraints lead to the same relations between the unknown coefficients as 
\bea
&&\{ C_{13} \to -128 + C_{1}/2, C_{16} \to 256 - C_{1} + (2 C_{10})/3 - C_{15},
C_{17} \to -(512/3) + C_{1} \nonumber \\ 
&&- (2 C_{10})/3,  C_{18} \to 256 - C_{1} - 2 C_{14} + 3 C_{15}, 
C_{19} \to 128 - C_{1}/4 + C_{12} - C_{14}/2,\nonumber \\ 
&& C_{2} \to C_{1}, 
C_{20} \to -128 + C_{1}/6 - (2 C_{12})/3 + C_{14}/3 + C_{15}/2, 
C_{21} \to C_{1}/3 - C_{10}/3 \nonumber \\ 
&&+ (2 C_{12})/3 - C_{14}/3 + C_{15}/2, 
C_{22} \to 128/3 - C_{1}/6 + C_{10}/8 - C_{15}/4, C_{23} \to 256/3 \nonumber \\ 
&&- C_{1}/3 + C_{10}/6, 
C_{24} \to 32 - C_{1}/8 + C_{10}/12 - C_{15}/8, C_{3} \to -256 - 4 C_{12}, 
C_{4} \to 128 \nonumber \\ 
&&- C_{1}/4 + C_{10}/2 + C_{11}/2, C_{5} \to -512 + 2 C_{1} - C_{10}, 
C_{6} \to 1024 - 4 C_{1} + 2 C_{10} + 8 C_{12},\nonumber \\ 
&& C_{7} \to 256 + 4 C_{12}, 
C_{8} \to -1280 + 4 C_{1} - 2 C_{10} - 4 C_{12}, C_{9} \to -256 + 2 C_{1} - C_{10} \}.
\eea

Having had these conditions, one can put them into the ansatz (\ref{FFRR}) to find the $ (\pa {F_4})^2 R^2 $ terms in 11 dimensions. Here also by doing so, we are left with a coupling with some determined and undetermined coefficients, but the terms containing undetermined coefficients vanish when we rewrite them in terms of independent variables. The final result is summarized as follows:
\bea
e^{-1}{\cal L}_{(\pa {F_4})^2 R^2}&=&\frac{32}{3} \lp 3 F^{eghi,f} F_{fghi,e} R_{abcd} R^{abcd} + 8 F^{b}{}_{fgh,i} F^{efgh,i} R_{abcd} R_{e}{}^{acd} \right. \nonumber \\ &&\qquad + 4 F_{fghi}{}^{,b} F^{fghi,e} R_{abcd} R_{e}{}^{acd} 
- 12 F^{bf}{}_{gh,i} F^{degh,i} R_{abcd} R_{e}{}^{a}{}_{f}{}^{c} \nonumber \\ 
&&\qquad+ 24 F^{bghi,d} F^{ef}{}_{gh,i} R_{abcd} R_{e}{}^{a}{}_{f}{}^{c}  - 16 F^{e}{}_{ghi}{}^{,b} F^{fghi,d} R_{abcd} R_{e}{}^{a}{}_{f}{}^{c} \nonumber \\ 
&&\qquad+ 24 F^{d}{}_{ghi}{}^{,b} F^{fghi,e} R_{abcd} R_{e}{}^{a}{}_{f}{}^{c}  + 12 F^{ae}{}_{gh,i} F^{bfgh,i} R_{abcd} R_{ef}{}^{cd} \nonumber \\ 
&&\qquad- 12 F^{e}{}_{ghi}{}^{,a} F^{fghi,b} R_{abcd} \
R_{ef}{}^{cd} - 24 F^{bghi,f} F^{ce}{}_{hi}{}^{,a} R_{abcd} R_{efg}{}^{d} \nonumber \\ 
&&\qquad + 24 F^{be}{}_{hi}{}^{,a} F^{cfhi,g} R_{abcd} R_{efg}{}^{d} - 120 F^{ce}{}_{hi}{}^{,a} F^{fghi,b} R_{abcd} \
R_{efg}{}^{d} \nonumber \\ 
&&\qquad+ 96 F^{be}{}_{hi}{}^{,a} F^{fghi,c} R_{abcd} R_{efg}{}^{d} - 48 F^{bc}{}_{hi}{}^{,a} F^{fghi,e} R_{abcd} R_{efg}{}^{d} \nonumber \\
&&\qquad \left. - 24 F^{acg}{}_{i}{}^{,e} F^{bdhi,f} R_{abcd} \
R_{efgh}  + 12 F^{cdgh}{}_{,i} F^{iabe,f} R_{abcd} R_{efgh}\rp.
\eea

\section{$ (\pa {F_4})^4 $ terms}\label{F4}

The basis for the $ (\pa {F_4})^4 $ terms (at linearised on-shell level) \cite{Peeters:2005tb} is given by 
\bea
&& C_1 F^{aehj,i} F_{bcde,a} F^{c}{}_{fgh}{}^{,b} \
F^{dg}{}_{ij}{}^{,f} + C_2 F^{a}{}_{fgh}{}^{,b} F_{bcde,a} \
F^{cdf}{}_{j,i} F^{eghi,j} \nonumber \\ 
&&+ C_3 F_{bcde,a} \
F^{b}{}_{fgh}{}^{,a} F^{cdf}{}_{j,i} F^{eghj,i}  + C_4 F^{ad}{}_{ij}{}^{,f} F_{bcde,a} \
F^{c}{}_{fgh}{}^{,b} F^{eghj,i} \nonumber \\ 
&&+ C_5 F_{bcde,a} \
F^{b}{}_{fgh}{}^{,a} F^{df}{}_{ij}{}^{,c} F^{eghj,i} + C_6 \
F^{ac}{}_{fg}{}^{,b} F_{bcde,a} F^{df}{}_{ij,h} F^{eghj,i}  \nonumber \\ 
&&+ C_7 F_{bcde,a} F^{bc}{}_{fg}{}^{,a} F^{df}{}_{ij,h} \
F^{eghj,i} + C_8 F_{bcde,a} F^{cdf}{}_{j}{}^{,b} F^{ehij,g} \
F_{fghi}{}^{,a} \nonumber \\ 
&& + C_9 F^{aghi,j} F_{bcde,a} \
F^{cde}{}_{j}{}^{,f} F_{fghi}{}^{,b} + C_{10} F^{aehi,j} F_{bcde,a} F^{cdg}{}_{j}{}^{,f} \
F_{fghi}{}^{,b} \nonumber \\ 
&& + C_{11} F^{acd}{}_{j}{}^{,f} F_{bcde,a} \
F^{ehij,g} F_{fghi}{}^{,b} + C_{12} F_{bcde,a} \
F^{b}{}_{fgh}{}^{,a} F^{cde}{}_{j,i} F^{fghi,j} \nonumber \\ 
&& + C_{13} F_{bcde,a} F^{cdej,i} F_{fghi}{}^{,a} \
F^{fgh}{}_{j}{}^{,b} + C_{14} F^{ac}{}_{fg}{}^{,b} \
F_{bcde,a} F^{de}{}_{ij,h} F^{fghj,i} \nonumber \\ 
&&+ C_{15} F_{bcde,a} \
F^{b}{}_{fgh}{}^{,a} F^{ehij,d} F^{fg}{}_{ij}{}^{,c}  + C_{16} F^{aeij,h} F_{bcde,a} F^{c}{}_{fgh}{}^{,b} \
F^{fg}{}_{ij}{}^{,d} \nonumber \\ 
&&+ C_{17} F^{ac}{}_{fg}{}^{,b} \
F_{bcde,a} F^{e}{}_{hij}{}^{,d} F^{fgij,h}  + C_{18} F_{bcde,a} F^{bc}{}_{fg}{}^{,a} F^{ehij,g} \
F^{f}{}_{hij}{}^{,d} \nonumber \\ 
&&+ C_{19} F_{bcde,a} \
F^{bcd}{}_{f}{}^{,a} F^{e}{}_{hij,g} F^{fhij,g}  + C_{20} F^{acde,j} F_{bcde,a} F_{fghi}{}^{,b} \
F^{ghi}{}_{j}{}^{,f} \nonumber \\ 
&&+ C_{21} F_{bcde,a} \
F^{b}{}_{fgh}{}^{,a} F^{cde}{}_{j,i} F^{ghij,f} + C_{22} \
F_{bcde,a} F^{bc}{}_{fg}{}^{,a} F^{e}{}_{hij}{}^{,d} F^{ghij,f} \nonumber \\ 
&& + C_{23} F_{bcde,a} F^{bcd}{}_{f}{}^{,a} F_{ghij}{}^{,e} \
F^{ghij,f} + C_{24} F_{bcde,a} F^{bcde,a} F_{ghij,f} \
F^{ghij,f}.\label{FFFF}
\eea

In order to determine the coefficients of these linear combinations of terms in the effective action, it is necessary to consider three following constraints: 
\begin{enumerate}
\item The terms in the form $ (\pa {F_4})^4 $ with no Killing index in $ D = 11 $ should transform to $ (\pa {F_4})^4 $ couplings in $ D = 10 $.
\item The terms with structure $ (\pa {F_4})^2 (\pa {F_4}_{z})^2 $ in $ D = 11 $ should convert to the terms $ (\pa {F_4})^2 (\pa H)^2 $ in $ D = 10 $.
\item The $ (\pa {F_4}_{z})^4 $ terms in $ D = 11 $ should produce $ (\pa H)^4 $ couplings in $ D = 10 $.
\end{enumerate}

Let us first focus on the terms with structure $ (\pa {F_4})^4 $ in $ 10 $ dimensions. To obtain the 10-dimensional version of these couplings, we first put the above basis under dimensional reduction and then select the terms $ (\pa {F_4})^4 $ in the dimensionally-reduced theory in which none of the 4-form field strengths contains any Killing index. These terms acquire the same form as 11-dimensional ones but with indices in ten dimensions using the transformation (\ref{4form}). The corresponding 10-dimensional couplings are also obtained in \cite{Policastro:2006vt,Bakhtiarizadeh:2015exa}. 

On the other hand, the $ (\pa {F_4})^2 (\pa H)^2 $ couplings in ten dimensions can be found by applying dimensional reduction on the above ansatz and choosing the terms in which two of the 4-form field strengths carry one Killing index while the two other ones carry no Killing index. These terms are then transformed to $ (\pa {F_4})^2 (\pa H)^2 $, using the compactification rule (\ref{4form}), and take the following explicit form:    
\bea
&& 8 C_{24} F_{efgh,i} F^{efgh,i} H_{abc,d} H^{abc,d} + 4 \
C_{23} F_{d}{}^{fgh,i} F_{efgh,i} H_{abc}{}^{,e} \
H^{abc,d} \nonumber \\ 
&&+ 6 C_{20} F_{e}{}^{fgh,i} F_{fghi,c} \
H_{abd}{}^{,e} H^{abc,d}  + 6 C_{19} F_{c}{}^{fgh,i} F_{efgh,i} \
H_{ab}{}^{e}{}_{,d} H^{abc,d} \nonumber \\ 
&&+ 3 C_{23} F_{fghi,e} \
F^{fghi}{}_{,c} H_{ab}{}^{e}{}_{,d} H^{abc,d}  + 3 C_{18} F_{cf}{}^{gh,i} F_{degh,i} H_{ab}{}^{e,f} \
H^{abc,d} \nonumber \\ 
&&- 6 C_{13} F_{c}{}^{ghi}{}_{,e} F_{dghi,f} \
H_{ab}{}^{e,f} H^{abc,d} + 3 C_{22} F_{cd}{}^{gh,i} \
F_{efgh,i} H_{ab}{}^{e,f} H^{abc,d} \nonumber \\ 
&& + 3 C_{21} F_{cd}{}^{gh,i} F_{eghi,f} H_{ab}{}^{e,f} \
H^{abc,d} - 6 C_9 F_{d}{}^{ghi}{}_{,e} F_{fghi,c} \
H_{ab}{}^{e,f} H^{abc,d} \nonumber \\ 
&&+ 6 C_{12} F_{c}{}^{ghi}{}_{,d} \
F_{fghi,e} H_{ab}{}^{e,f} H^{abc,d}  - 2 C_{11} F_{bf}{}^{gh,i} F_{cghi,e} H_{ad}{}^{e,f} \
H^{abc,d} \nonumber \\ 
&&+ 4 C_8 F_{be}{}^{gh,i} F_{cghi,f} \
H_{ad}{}^{e,f} H^{abc,d} + 2 C_{17} F_{bc}{}^{gh,i} \
F_{efgi,h} H_{ad}{}^{e,f} H^{abc,d} \nonumber \\ 
&& - 2 C_{21} F_{b}{}^{ghi}{}_{,c} F_{eghi,f} \
H_{ad}{}^{e,f} H^{abc,d} - 2 C_{11} F_{be}{}^{gh,i} \
F_{fghi,c} H_{ad}{}^{e,f} H^{abc,d} \nonumber \\ 
&&+ 2 C_7 \
F_{be}{}^{gh,i} F_{cfgi,h} H_{a}{}^{ef}{}_{,d} H^{abc,d} 
+ 2 C_{22} F_{b}{}^{ghi}{}_{,c} F_{eghi,f} \
H_{a}{}^{ef}{}_{,d} H^{abc,d} \nonumber \\ 
&&+ 2 C_{18} \
F_{b}{}^{ghi}{}_{,e} F_{fghi,c} H_{a}{}^{ef}{}_{,d} H^{abc,d}  + 2 C_5 F_{bde}{}^{h,i} F_{cfhi,g} H_{a}{}^{ef,g} \
H^{abc,d} \nonumber \\ 
&&+ 4 C_3 F_{be}{}^{hi}{}_{,d} F_{cfhi,g} \
H_{a}{}^{ef,g} H^{abc,d} - 2 C_{16} F_{bcg}{}^{h,i} \
F_{dehi,f} H_{a}{}^{ef,g} H^{abc,d} \nonumber \\ 
&& + 2 C_{10} F_{bg}{}^{hi}{}_{,c} F_{dehi,f} \
H_{a}{}^{ef,g} H^{abc,d} + 2 C_{15} F_{bce}{}^{h,i} \
F_{dfgh,i} H_{a}{}^{ef,g} H^{abc,d} \nonumber \\ 
&& - 2 C_{16} F_{bc}{}^{hi}{}_{,e} F_{dfgh,i} \
H_{a}{}^{ef,g} H^{abc,d} + 4 C_2 F_{bg}{}^{hi}{}_{,e} \
F_{dfhi,c} H_{a}{}^{ef,g} H^{abc,d} \nonumber \\ 
&&+ 2 C_{15} \
F_{bce}{}^{h,i} F_{dfhi,g} H_{a}{}^{ef,g} H^{abc,d}  + 2 C_{17} F_{bcd}{}^{h,i} F_{eghi,f} H_{a}{}^{ef,g} \
H^{abc,d} \nonumber \\ 
&&+ 4 C_{14} F_{bd}{}^{hi}{}_{,c} F_{eghi,f} \
H_{a}{}^{ef,g} H^{abc,d} + 2 C_4 F_{bde}{}^{h,i} \
F_{fghi,c} H_{a}{}^{ef,g} H^{abc,d} \nonumber \\ 
&& + 2 C_{10} F_{bd}{}^{hi}{}_{,e} F_{fghi,c} \
H_{a}{}^{ef,g} H^{abc,d} + C_7 F_{be}{}^{gh,i} \
F_{cfgh,i} H^{abc,d} H_{d}{}^{ef}{}_{,a} \nonumber \\ 
&&+ 2 C_6 \
F_{be}{}^{gh,i} F_{cfgi,h} H^{abc,d} H_{d}{}^{ef}{}_{,a}  + 2 C_{14} F_{bc}{}^{gh,i} F_{efgi,h} H^{abc,d} \
H_{d}{}^{ef}{}_{,a} \nonumber \\ 
&&-  C_{17} F_{bc}{}^{gh,i} \
F_{eghi,f} H^{abc,d} H_{d}{}^{ef}{}_{,a}  + 4 C_1 F_{aeg}{}^{h,i} F_{bfhi,c} H^{abc,d} \
H_{d}{}^{ef,g} \nonumber \\ 
&&-  C_4 F_{ae}{}^{hi}{}_{,f} F_{bghi,c} \
H^{abc,d} H_{d}{}^{ef,g} + C_4 F_{abg}{}^{h,i} \
F_{cehi,f} H^{abc,d} H_{d}{}^{ef,g} \nonumber \\ 
&& + C_5 F_{abe}{}^{h,i} F_{cfgh,i} H^{abc,d} \
H_{d}{}^{ef,g} -  C_{15} F_{ab}{}^{hi}{}_{,e} \
F_{cfhi,g} H^{abc,d} H_{d}{}^{ef,g} \nonumber \\ 
&&-  C_5 \
F_{ae}{}^{hi}{}_{,b} F_{cfhi,g} H^{abc,d} H_{d}{}^{ef,g}  + C_{21} F_{abc}{}^{h,i} F_{efgh,i} H^{abc,d} \
H_{d}{}^{ef,g} \nonumber \\ 
&&+ 2 C_{10} F_{abg}{}^{h,i} F_{efhi,c} \
H^{abc,d} H_{d}{}^{ef,g} -  C_{16} F_{ag}{}^{hi}{}_{,b} \
F_{efhi,c} H^{abc,d} H_{d}{}^{ef,g} \nonumber \\ 
&& - 2 C_{22} F_{abc}{}^{h,i} F_{efhi,g} H^{abc,d} \
H_{d}{}^{ef,g} -  C_{17} F_{ab}{}^{hi}{}_{,c} \
F_{eghi,f} H^{abc,d} H_{d}{}^{ef,g} \nonumber \\ 
&&+ C_4 \
F_{abe}{}^{h,i} F_{fghi,c} H^{abc,d} H_{d}{}^{ef,g}  + C_{16} F_{ab}{}^{hi}{}_{,e} F_{fghi,c} H^{abc,d} \
H_{d}{}^{ef,g} \nonumber \\ 
&&+ 2 C_3 F_{abe}{}^{h,i} F_{cfgh,i} \
H^{abc,d} H^{efg}{}_{,d} -  C_5 F_{abe}{}^{h,i} \
F_{cfhi,g} H^{abc,d} H^{efg}{}_{,d} \nonumber \\ 
&& + C_{15} F_{ab}{}^{hi}{}_{,e} F_{cfhi,g} H^{abc,d} \
H^{efg}{}_{,d} + C_{12} F_{abc}{}^{h,i} F_{efgi,h} \
H^{abc,d} H^{efg}{}_{,d} \nonumber \\ 
&&-  C_{21} F_{abc}{}^{h,i} \
F_{efhi,g} H^{abc,d} H^{efg}{}_{,d}  - 2 C_1 F_{ade}{}^{i}{}_{,f} F_{bghi,c} H^{abc,d} \
H^{efg,h} \nonumber \\ 
&&-  C_5 F_{abe}{}^{i}{}_{,h} F_{cdfi,g} \
H^{abc,d} H^{efg,h} + C_4 F_{abe}{}^{i}{}_{,f} \
F_{cdgh,i} H^{abc,d} H^{efg,h} \nonumber \\ 
&& -  C_{15} F_{abef}{}^{,i} F_{cdgi,h} H^{abc,d} \
H^{efg,h} + C_{11} F_{abdh}{}^{,i} F_{cefi,g} H^{abc,d} \
H^{efg,h} \nonumber \\ 
&&- 2 C_8 F_{abd}{}^{i}{}_{,h} F_{cefi,g} \
H^{abc,d} H^{efg,h}  + C_8 F_{abde}{}^{,i} F_{cfgh,i} H^{abc,d} H^{efg,h} \
\nonumber \\ 
&&+ 2 C_3 F_{abe}{}^{i}{}_{,h} F_{cfgi,d} H^{abc,d} \
H^{efg,h} + 4 C_7 F_{abe}{}^{i}{}_{,d} F_{cfhi,g} \
H^{abc,d} H^{efg,h} \nonumber \\ 
&& + 4 C_6 F_{ade}{}^{i}{}_{,b} F_{cfhi,g} H^{abc,d} \
H^{efg,h} -  C_5 F_{abe}{}^{i}{}_{,f} F_{cghi,d} \
H^{abc,d} H^{efg,h} \nonumber \\ 
&&+ C_{13} F_{abch}{}^{,i} F_{defg,i} \
H^{abc,d} H^{efg,h}  + 2 C_2 F_{abh}{}^{i}{}_{,e} F_{dfgi,c} H^{abc,d} \
H^{efg,h} \nonumber \\ 
&&- 2 C_{18} F_{abc}{}^{i}{}_{,e} F_{dfgi,h} \
H^{abc,d} H^{efg,h} + 2 C_{19} F_{abc}{}^{i}{}_{,d} \
F_{efgi,h} H^{abc,d} H^{efg,h}  \nonumber \\ 
&&+ C_{11} F_{abde}{}^{,i} F_{fghi,c} H^{abc,d} \
H^{efg,h}.
\eea

As already mentioned above, the corresponding 10-dimensional couplings in type IIA supergravity can be found in \cite{Policastro:2006vt,Garousi:2013nfw,Bakhtiarizadeh:2017bpl,Bakhtiarizadeh:2013zia}. 
Imposing the first two constraints with this requirement that the couplings derived from dimensional reduction on a circle should be consistent with the corresponding $ 10 $-dimensional ones, leads to the same relations between the unknown coefficients as  

\bea
&&\{C_{11} \to 0, 
C_{16} \to -128 + C_{10} - 2 C_{14}, C_{19} \to 64/9 + C_{14}/9, 
C_{2} \to 128 - C_{10} + 3 C_{14},\nonumber\\&& 
C_{20} \to -(128/9) + 2 C_{12} - C_{13} - C_{14}/9 + C_{18}/3, 
C_{21} \to 128 - (2 C_{10})/3 - 3 C_{12} \nonumber\\&& + 3 C_{13} + (4 C_{14})/3 - C_{18}, 
C_{22} \to -(448/3) + (2 C_{10})/3 + 3 C_{12} - 3 C_{13} - 2 C_{14} + C_{18}, \nonumber\\&& 
C_{23} \to -(32/9) - C_{13}/4 - C_{14}/18 + C_{18}/12, 
C_{24} \to -(2/9) + C_{12}/32, C_{3} \to -64 + C_{10} \nonumber\\&&- 2 C_{14},  
C_{5} \to 512 - 8 C_{10} + 16 C_{14} - 2 C_{15} + C_{4}, 
C_{6} \to -768 + C_{1}/2 + 6 C_{10} + 18 C_{12} \nonumber\\&&- 18 C_{13} - 16 C_{14}  - 2 C_{17}, 
C_{7} \to 192 - 9 C_{12} + 9 C_{13} + 2 C_{14} + C_{15} + C_{17} - C_{4}/2, \nonumber\\&& 
C_{8} \to 128 + 2 C_{14} - 3 C_{18}, 
C_{9} \to -(128/9) + C_{12} - C_{13}\}. \label{condi1}
\eea

Now, we are going to impose the third constraint. To this end, among other couplings in dimensionally reduced theory, we select the terms in which each 4-form field strengths carries one Killing index. They convert to $ (\pa H)^4 $ in the compactified theory due to the transformation (\ref{4form}). They read:
\bea
&&-2 C_5 H_{ad}{}^{e,f} H^{abc,d} H_{be}{}^{g,h} H_{cfg,h} \
+ 4 C_3 H_{a}{}^{ef}{}_{,d} H^{abc,d} H_{be}{}^{g,h} \
H_{cfg,h} \nonumber \\ 
&&+ 2 C_7 H_{a}{}^{ef}{}_{,d} H^{abc,d} \
H_{be}{}^{g,h} H_{cfh,g}  - 4 C_8 H_{ad}{}^{e,f} H^{abc,d} H_{be}{}^{g,h} \
H_{cgh,f} \nonumber \\ 
&&+ 6 C_{18} H_{ab}{}^{e,f} H^{abc,d} \
H_{cf}{}^{g,h} H_{deg,h} + 2 C_{15} H_{ab}{}^{e,f} \
H^{abc,d} H_{ce}{}^{g,h} H_{dfg,h}  \nonumber \\ 
&&- 2 C_{16} H_{ab}{}^{e,f} H^{abc,d} H_{c}{}^{gh}{}_{,e} \
H_{dfg,h} + (4 C_2 + C_6) H_{a}{}^{ef,g} \
H^{abc,d} H_{bg}{}^{h}{}_{,e} H_{dfh,c} \nonumber \\ 
&&+ (2 C_{16} + C_4) H_{ab}{}^{e,f} H^{abc,d} H_{cf}{}^{g,h} \
H_{dgh,e} + 2 C_{15} H_{ab}{}^{e,f} H^{abc,d} \
H_{ce}{}^{g,h} H_{dgh,f} \nonumber \\ 
&&- 9 C_{13} H_{ab}{}^{e,f} \
H^{abc,d} H_{c}{}^{gh}{}_{,e} H_{dgh,f}  -  C_1 H_{ad}{}^{e,f} H^{abc,d} H_{b}{}^{gh}{}_{,c} \
H_{efg,h} \nonumber \\ 
&&+ 6 C_{22} H_{ab}{}^{e,f} H^{abc,d} \
H_{cd}{}^{g,h} H_{efg,h} + 9 C_{19} H_{ab}{}^{e}{}_{,d} \
H^{abc,d} H_{c}{}^{fg,h} H_{efg,h} \nonumber \\ 
&& + (- C_{18} + C_{22} + 12 C_{23}) \
H_{abc}{}^{,e} H^{abc,d} H_{d}{}^{fg,h} H_{efg,h} - 2 C_{17} H_{abd}{}^{,e} H^{abc,d} H_{c}{}^{fg,h} H_{efh,g} \nonumber \
\\ 
&& + 2 C_7 H_{ab}{}^{e}{}_{,d} H^{abc,d} H_{c}{}^{fg,h} \
H_{efh,g} + 2 (2 C_{14} + C_6) \
H_{ad}{}^{e}{}_{,b} H^{abc,d} H_{c}{}^{fg,h} H_{efh,g} \nonumber \\ 
&&- 2 (C_1 + 2 C_{10}) H_{ad}{}^{e,f} H^{abc,d} H_{bf}{}^{g,h} \
H_{egh,c} -  C_5 H_{ab}{}^{e,f} H^{abc,d} H_{cf}{}^{g,h} \
H_{egh,d} \nonumber \\ 
&&+ C_3 H_{ab}{}^{e,f} H^{abc,d} \
H_{c}{}^{gh}{}_{,f} H_{egh,d}  + (-6 C_{21} -  C_5) H_{ab}{}^{e,f} H^{abc,d} \
H_{cd}{}^{g,h} H_{egh,f} \nonumber \\ 
&&+ C_3 H_{ab}{}^{e,f} H^{abc,d} \
H_{c}{}^{gh}{}_{,d} H_{egh,f} + (C_{19} + 16 C_{24}) H_{abc,d} H^{abc,d} H_{efg,h} H^{efg,h} \nonumber \\ 
&&+ \
2 (2 C_{11} + C_4) H_{ad}{}^{e,f} H^{abc,d} \
H_{be}{}^{g,h} H_{fgh,c} + (- C_{10} - 9 C_9) H_{ab}{}^{e,f} H^{abc,d} H_{d}{}^{gh}{}_{,e} \
H_{fgh,c} \nonumber \\ 
&&- 9 C_{20} H_{abd}{}^{,e} H^{abc,d} \
H_{e}{}^{fg,h} H_{fgh,c} + 2 C_{17} H_{ab}{}^{e,f} \
H^{abc,d} H_{cd}{}^{g,h} H_{fgh,e} \nonumber \\ 
&& -  C_8 H_{abd}{}^{,e} H^{abc,d} H_{c}{}^{fg,h} \
H_{fgh,e} + 9 C_{12} H_{ab}{}^{e,f} H^{abc,d} \
H_{c}{}^{gh}{}_{,d} H_{fgh,e} \nonumber \\ 
&&+ C_{21} H_{abc}{}^{,e} \
H^{abc,d} H_{d}{}^{fg,h} H_{fgh,e}  + (C_{14} + C_2) H_{ab}{}^{e,f} H^{abc,d} \
H_{d}{}^{gh}{}_{,c} H_{fgh,e}.
\eea

The associated couplings in type II theories can be obtained from the known $ R^4 $ action by substituting the Riemann curvature by the generalized Riemann curvature \cite{Gross:1986mw}. Matching these couplings leads to the following relations between the unknown coefficients
\bea
&&\{C_{22} \to 
32 - C_{10}/9 - (2 C_{11})/9 - 3 C_{12} + 2 C_{13} - C_{14}/2 + (2 C_{16})/9 - (
2 C_{18})/3 \nonumber\\&&- (3 C_{19})/2 + C_{2}/3 + C_{20} - (4 C_{21})/3, 
C_{23} \to -(64/9) + C_{11}/18 + C_{12}/4 - C_{13}/2 \nonumber\\&&+ C_{14}/18 - C_{16}/18 + C_{18}/
6 - C_{2}/9 + C_{21}/12, C_{24} \to 1/9 + C_{12}/32 + C_{14}/192 \nonumber\\&&- (3 C_{19})/64,
C_{3} \to 32 + C_{14}/2 + (9 C_{19})/2 - C_{2}, 
C_{5} \to -384 + 4 C_{11} - 2 C_{14} - 2 C_{15} \nonumber\\&&- 4 C_{16} - 18 C_{19} + 4 C_{2} + C_{4},
C_{6} \to -128 + C_{1}/2 + 2 C_{11} + 18 C_{12} - 18 C_{13} + 2 C_{14} \nonumber\\&&- 2 C_{16} - 
2 C_{17} + 18 C_{19} - 8 C_{2}, 
C_{7} \to 128 + C_{10}/3 - (7 C_{11})/3 + 3 C_{13} + C_{15} + (7 C_{16})/3 \nonumber\\&&+ C_{17} + 
2 C_{18} - 9 C_{19} + 2 C_{2} - 3 C_{20} + C_{21} - C_{4}/2, 
C_{8} \to -96 - C_{10}/3 + (4 C_{11})/3 \nonumber\\&&+ 18 C_{12} - 12 C_{13} + (7 C_{14})/2 - (
4 C_{16})/3 + C_{18} + (27 C_{19})/2 - 3 C_{2} - 6 C_{20} + 2 C_{21},\nonumber\\&& 
C_{9} \to -(64/9) - C_{10}/9 + C_{12} - C_{13} + (2 C_{14})/9 + C_{19} - C_{2}/9\}.\label{condi2}
\eea
By substituting the conditions (\ref{condi1}) into the basis (\ref{FFFF}), one finds the following couplings between four $ 4 $-form field strengths in 11 dimensions:
\bea
e^{-1}{\cal L}_{(\pa {F_4})^4}&=&\frac{2}{9} \lp 576 F^{a}{}_{fgh}{}^{,b} F_{bcde,a} F^{cdf}{}_{j,i} \
F^{eghi,j} - 288 F_{bcde,a} F^{b}{}_{fgh}{}^{,a} F^{cdf}{}_{j,i} \
F^{eghj,i} \right.\nonumber \\ 
&&\quad+ 2304 F_{bcde,a} F^{b}{}_{fgh}{}^{,a} \
F^{df}{}_{ij}{}^{,c} F^{eghj,i}  - 3456 F^{ac}{}_{fg}{}^{,b} F_{bcde,a} F^{df}{}_{ij,h} F^{eghj,i} \
\nonumber \\ 
&&\quad+ 864 F_{bcde,a} F^{bc}{}_{fg}{}^{,a} F^{df}{}_{ij,h} F^{eghj,i} + \
576 F_{bcde,a} F^{cdf}{}_{j}{}^{,b} F^{ehij,g} F_{fghi}{}^{,a} \nonumber \\ 
&&\quad- 64 F^{aghi,j} F_{bcde,a} F^{cde}{}_{j}{}^{,f} F_{fghi}{}^{,b} - \
576 F^{aeij,h} F_{bcde,a} F^{c}{}_{fgh}{}^{,b} F^{fg}{}_{ij}{}^{,d} \nonumber \\ 
&&\quad+ \
32 F_{bcde,a} F^{bcd}{}_{f}{}^{,a} F^{e}{}_{hij,g} F^{fhij,g}  - 64 F^{acde,j} F_{bcde,a} F_{fghi}{}^{,b} F^{ghi}{}_{j}{}^{,f} \nonumber \\ 
&&\quad+ \
576 F_{bcde,a} F^{b}{}_{fgh}{}^{,a} F^{cde}{}_{j,i} F^{ghij,f} - 672 
F_{bcde,a} F^{bc}{}_{fg}{}^{,a} F^{e}{}_{hij}{}^{,d} F^{ghij,f} \nonumber \\ 
&&\quad \left. - 16 F_{bcde,a} F^{bcd}{}_{f}{}^{,a} F_{ghij}{}^{,e} F^{ghij,f} -  \
F_{bcde,a} F^{bcde,a} F_{ghij,f} F^{ghij,f}\rp,
\eea
plus some other terms with unknown coefficients which are zero for the reasons already mentioned. This coupling which has been obtained from the above constraints $ 1 $ and $ 2 $, automatically satisfies the constraint $ 3 $. But the conditions (\ref{condi2}), which are obtained by applying the constraint $ 3 $, do not fix all the unknown coefficients and consequently lead to an incorrect coupling that does not satisfy the other two constraints. This indicates that each of the above constraints alone is necessary but not sufficient to obtain the correct coupling.   
    
\section{Discussion}\label{dis}

In this paper we have presented a systematic derivation of the modifications to the eleven-dimensional supergravity. In contrast to existing approaches, our analysis is based on the dimensional reduction of eleven-dimensional supergravity. Given the complexity of higher-derivative supergravity actions, it is most encouraging that the use of dimensional reduction information has enabled us to find these corrections.

One may also use the algorithm introduced in \cite{Bakhtiarizadeh:2015exa} to reduce the tensor polynomials and rewrite the couplings in their minimal-term forms. We observe that the coupling (\ref{RRRR}) is in its minimal-term form. On the other hand, the reduced (relativity normalized) form of the $ (\pa {F_4})^2 R^2 $ terms can be written as
\bea
e^{-1}{\cal L}_{(\pa {F_4})^2 R^2}&=&\frac{64}{3}  \lp 12 F^{a g h}{}_{i}{}^{,e} F^{b d f i ,c} R_{a b c d} R_{e f g h}  + 12 F^{a c g}{}_{i}{}^{,e} F^{b d f i ,h} R_{a b c d} R_{e f g h} \right. \nonumber\\&&\qquad+ 12 F^{b g h i ,f} F^{c e}{}_{h i}{}^{,a} R_{a b c d} R_{e f g}{}^{d} - 3 F^{b a h i ,g} F^{e f}{}_{h i}{}^{,c} R_{a b c d} R_{e f g}{}^{d}\nonumber\\&&\qquad+ 2 F^{b g h i ,f} F^{e}{}_{g h i}{}^{,a} R_{a b c d} R_{e f}{}^{c d} - 2 F^{e}{}_{g h i}{}^{,a} F^{f g h i ,b} R_{a b c d} R_{e f}{}^{c d} \nonumber\\&&\qquad - 24 F^{b e}{}_{h i}{}^{,a} F^{f g h i ,c} R_{a b c d} R_{e f g}{}^{d} + 4 F^{e}{}_{g h i}{}^{,b} F^{f g h i ,d} R_{a b c d} R_{e}{}^{a}{}_{f}{}^{c} \nonumber\\&&\qquad \left.+ 3 F^{c d g h}{}_{,i} F^{i a b e ,f} R_{a b c d} R_{e f g h} \rp. \label{F4F4RR}
\eea
In the other words, they are different presentations that are equivalent up to symmetries of the various tensors. Furthermore, the  (relativity normalized) $ (\pa {F_4})^4 $ terms are given by the following economical form
\bea
e^{-1}{\cal L}_{(\pa {F_4})^4}&=& -\frac{128}{3} \lp 72 F_{a b f g ,e} F^{a b c d ,e} F_{c d i j ,h} F^{f g i j ,h}-36 F_{a b f g ,i} F^{a b c d ,e} F_{c d j h ,i} F^{f g j h ,e} \right.\nonumber\\&&\qquad\quad-64 F_{a b c f ,g} F^{a b c d ,e} F_{d i j h ,e} F^{f i j h ,g} -F_{a b c d ,e} F^{a b c d ,e} F_{f g i j ,h} F^{f g i j ,h} \nonumber\\&&\qquad\quad\left.+6 F_{a b c d ,f} F^{a b c d ,e} F_{g i j h ,f} F^{g i j h ,e} \rp.\label{F4F4F4F4}
\eea

Our findings in the present paper agree with the results that have been obtained in \cite{Peeters:2005tb} using superparticle vertex operator correlators in the light-cone gauge, up to an overall factor. We also check our results by calculating the scattering amplitude of massless states in 11 dimensions and find an exact agreement.

As a next future work, it is also interesting to consider higher-derivative corrections to supergravity in twelve dimensions \cite{Khviengia:1997rh,Choi:2014vya,Choi:2015gia,Berman:2015rcc,Minasian:2015bxa}, whose dimensional reduction on a circle and on a torus yields 11-dimensional and type IIB supergravity, respectively. This also provides the effective field theory of F-theory \cite{Vafa:1996xn}. Applications to black hole physics \cite{Hanada:2008ez}, brane solutions \cite{Lu:1995cs,Lu:1995yn,Gueven:1992hh} and cosmology \cite{Ciupke:2016vkc} are also important directions.


\section*{Acknowledgement}\addcontentsline{toc}{section}{Acknowledgement}

This work has been financially supported by the research deputy of Sirjan University of Technology.


\providecommand{\href}[2]{#2}\begingroup\raggedright
\endgroup
\end{document}